\documentclass[a4paper]{article} 

\let\originalleft\left
\let\originalright\right
\renewcommand{\left}{\mathopen{}\mathclose\bgroup\originalleft}
\renewcommand{\right}{\aftergroup\egroup\originalright}

\DeclareSymbolFont{matha}{OML}{txmi}{m}{it}
\DeclareMathSymbol{\varv}{\mathord}{matha}{118}

\ifdefined\myframe
\renewenvironment{myframe}[2]{\section{#1}\begin{frame}{#2}\vspace{-10pt}}{\end{frame}} 
\else
\newenvironment{myframe}[2]{\section{#1}\begin{frame}{#2}\vspace{-10pt}}{\end{frame}}
\fi

\ifdefined\U
\renewcommand{\U}[1]{\underline{#1}}
\else
\newcommand{\U}[1]{\underline{#1}}
\fi

\ifdefined\UU
\renewcommand{\UU}[1]{\underline{\underline{#1}}}
\else
\newcommand{\UU}[1]{\underline{\underline{#1}}}
\fi

\ifdefined\mybullet
\renewcommand{\mybullet}{\vspace{2mm}\\$\bullet$ }
\else
\newcommand{\mybullet}{\vspace{2mm}\\$\bullet$ }
\fi

\ifdefined\mybulletEQ
\renewcommand{\mybulletEQ}[1]{$\bullet$ {\bf #1}\vspace{1mm}\\}
\else
\newcommand{\mybulletEQ}[1]{$\bullet$ {\bf #1}\vspace{1mm}\\}
\fi

\ifdefined\fract
\renewcommand{\fract}[2]{{\textstyle \frac{#1}{#2}}}
\else
\newcommand{\fract}[2]{{\textstyle \frac{#1}{#2}}}
\fi

\ifdefined\rect
\renewcommand{\fract}[1]{{\textstyle \frac{1}{#1}}}
\else
\newcommand{\rect}[1]{{\textstyle \frac{1}{#1}}}
\fi

\ifdefined\fracd
\renewcommand{\fracd}[2]{\frac{\displaystyle{#1}}{\displaystyle{#2}}}
\else
\newcommand{\fracd}[2]{\frac{\displaystyle{#1}}{\displaystyle{#2}}}
\fi

\ifdefined\recd
\renewcommand{\recd}[1]{\frac{\displaystyle 1}{\displaystyle{#1}}}
\else
\newcommand{\recd}[1]{\frac{\displaystyle 1}{\displaystyle{#1}}}
\fi

\ifdefined\pdd
\renewcommand{\pdd}[2]{\frac{\displaystyle{\partial{#1}}}{\displaystyle{\partial{#2}}}}
\else
\newcommand{\pdd}[2]{\frac{\displaystyle{\partial{#1}}}{\displaystyle{\partial{#2}}}}
\fi

\ifdefined\biggg
\renewcommand{\biggg}[1]{\scalebox{1.2}{\Bigg{#1}}}
\else
\newcommand{\biggg}[1]{\scalebox{1.2}{\Bigg{#1}}}
\fi

\ifdefined\Biggg
\renewcommand{\Biggg}[1]{\scalebox{1.4}{\Bigg{#1}}}
\else
\newcommand{\Biggg}[1]{\scalebox{1.4}{\Bigg{#1}}}
\fi

\ifdefined\Re
\renewcommand{\Re}{\operatorname{Re}}
\else
\newcommand{\Re}{\operatorname{Re}}
\fi

\ifdefined\Im
\renewcommand{\Im}{\operatorname{Im}}
\else
\newcommand{\Im}{\operatorname{Im}}
\fi

\ifdefined\Arch
\renewcommand{\Arch}{\operatorname{Ar\,ch}}
\else
\newcommand{\Arch}{\operatorname{Ar\,ch}}
\fi

\ifdefined\Arsh
\renewcommand{\Arsh}{\operatorname{Ar\,sh}}
\else
\newcommand{\Arsh}{\operatorname{Ar\,sh}}
\fi

\ifdefined\Arth
\renewcommand{\Arth}{\operatorname{Arth}}
\else
\newcommand{\Arth}{\operatorname{Arth}}
\fi

\ifdefined\ch
\renewcommand{\ch}{\operatorname{ch}}
\else
\newcommand{\ch}{\operatorname{ch}}
\fi

\ifdefined\sh
\renewcommand{\sh}{\operatorname{sh}}
\else
\newcommand{\sh}{\operatorname{sh}}
\fi

\ifdefined\th
\renewcommand{\th}{\operatorname{th}}
\else
\newcommand{\th}{\operatorname{th}}
\fi

\ifdefined\Ln
\renewcommand{\Ln}{\operatorname{Ln}}
\else
\newcommand{\Ln}{\operatorname{Ln}}
\fi

\ifdefined\tg
\renewcommand{\tg}{\operatorname{tg}}
\else
\newcommand{\tg}{\operatorname{tg}}
\fi

\ifdefined\ctg
\renewcommand{\ctg}{\operatorname{ctg}}
\else
\newcommand{\ctg}{\operatorname{ctg}}
\fi

\ifdefined\intl
\renewcommand{\intl}{\int\limits}
\else
\newcommand{\intl}{\int\limits}
\fi

\ifdefined\ointl
\renewcommand{\ointl}{\oint\limits}
\else
\newcommand{\ointl}{\oint\limits}
\fi

\ifdefined\integrated
\renewcommand{\integrated}[3]{\left\{{#1}\right\}\left.\vphantom{#1}\right|_{#2}^{#3}}
\else
\newcommand{\integrated}[3]{\left\{{#1}\right\}\left.\vphantom{#1}\right|_{#2}^{#3}}
\fi

\ifdefined\pd
\renewcommand{\pd}[2]{\frac{\partial{#1}}{\partial{#2}}}
\else
\newcommand{\pd}[2]{\frac{\partial{#1}}{\partial{#2}}}
\fi

\ifdefined\rec
\renewcommand{\rec}[1]{\frac{1}{#1}}
\else
\newcommand{\rec}[1]{\frac{1}{#1}}
\fi

\ifdefined\gvec
\renewcommand{\gvec}[1]{\mbox{\boldmath${#1}$}}
\else
\newcommand{\gvec}[1]{\mbox{\boldmath${#1}$}}
\fi

\ifdefined\cvec
\renewcommand{\cvec}[1]{\mbox{\boldmath${#1}$}}
\else
\newcommand{\cvec}[1]{\mbox{\boldmath${#1}$}}
\fi

\ifdefined\td
\renewcommand{\td}[2]{\frac{d{#1}}{d{#2}}}
\else
\newcommand{\td}[2]{\frac{d{#1}}{d{#2}}}
\fi

\ifdefined\md
\renewcommand{\md}[2]{\frac{\mathrm{d}{#1}}{\mathrm{d}{#2}}}
\else
\newcommand{\md}[2]{\frac{\mathrm{d}{#1}}{\mathrm{d}{#2}}}
\fi

\ifdefined\z
\renewcommand{\z}[1]{\left({#1}\right)}
\else
\newcommand{\z}[1]{\left({#1}\right)}
\fi

\ifdefined\ae
\renewcommand{\ae}[1]{\left|{#1}\right|}
\else
\newcommand{\ae}[1]{\left|{#1}\right|}
\fi

\ifdefined\sz
\renewcommand{\sz}[1]{\left[{#1}\right]}
\else
\newcommand{\sz}[1]{\left[{#1}\right]}
\fi

\ifdefined\kz
\renewcommand{\kz}[1]{\left\{{#1}\right\}}
\else
\newcommand{\kz}[1]{\left\{{#1}\right\}}
\fi

\ifdefined\B
\renewcommand{\B}[1]{\mathbb{#1}}
\else
\newcommand{\B}[1]{\mathbb{#1}}
\fi

\ifdefined\m
\renewcommand{\m}[1]{\mathrm{#1}}
\else
\newcommand{\m}[1]{\mathrm{#1}}
\fi

\ifdefined\tn
\renewcommand{\tn}[1]{\textnormal{#1}}
\else
\newcommand{\tn}[1]{\textnormal{#1}}
\fi

\ifdefined\o
\renewcommand{\o}[1]{\operatorname{#1}}
\else
\newcommand{\o}[1]{\operatorname{#1}}
\fi

\ifdefined\c
\renewcommand{\c}[1]{\mathcal{#1}}
\else
\newcommand{\c}[1]{\mathcal{#1}}
\fi

\ifdefined\v
\renewcommand{\v}[1]{\mathbf{#1}}
\else
\newcommand{\v}[1]{\mathbf{#1}}
\fi

\ifdefined\Eq
\renewcommand{\Eq}[1]{Eq.~(\ref{#1})}
\else
\newcommand{\Eq}[1]{Eq.~(\ref{#1})}
\fi

\ifdefined\Eqs
\renewcommand{\Eqs}[2]{Eqs.~(\ref{#1}) and (\ref{#2})}
\else
\newcommand{\Eqs}[2]{Eqs.~(\ref{#1}) and (\ref{#2})}
\fi

\ifdefined\a
\renewcommand{\a}[1]{\aref({#1})}
\else
\newcommand{\a}[1]{\aref({#1})}
\fi

\ifdefined\A
\renewcommand{\A}[1]{\Aref({#1})}
\else
\newcommand{\A}[1]{\Aref({#1})}
\fi

\ifdefined\r

\renewcommand{\r}[1]{(\ref{#1})}
\else
\newcommand{\r}[1]{(\ref{#1})}
\fi

\ifdefined\comm
\renewcommand{\comm}[2]{\left[{#1},{#2}\right]}
\else
\newcommand{\comm}[2]{\left[{#1},{#2}\right]}
\fi

\ifdefined\follows
\renewcommand{\follows}{\quad\Rightarrow\quad}
\else
\newcommand{\follows}{\quad\Rightarrow\quad}
\fi

\ifdefined\Follows
\renewcommand{\Follows}{\qquad\Rightarrow\qquad}
\else
\newcommand{\Follows}{\qquad\Rightarrow\qquad}
\fi

\ifdefined\followse
\renewcommand{\followse}{\quad\Rightarrow}
\else
\newcommand{\followse}{\quad\Rightarrow}
\fi

\ifdefined\bfollows
\renewcommand{\bfollows}{\Rightarrow\quad}
\else
\newcommand{\bfollows}{\Rightarrow\quad}
\fi

\ifdefined\equivalent
\renewcommand{\equivalent}{\quad\Leftrightarrow\quad}
\else
\newcommand{\equivalent}{\quad\Leftrightarrow\quad}
\fi

\ifdefined\obs
\renewcommand{\obs}[1]{\left\langle{#1}\right\rangle}
\else
\newcommand{\obs}[1]{\left\langle{#1}\right\rangle}
\fi

\ifdefined\ket
\renewcommand{\ket}[1]{\left|{#1}\right\rangle}
\else
\newcommand{\ket}[1]{\left|{#1}\right\rangle}
\fi

\ifdefined\bra
\renewcommand{\bra}[1]{\left\langle{#1}\right|}
\else
\newcommand{\bra}[1]{\left\langle{#1}\right|}
\fi

\ifdefined\braket
\renewcommand{\braket}[2]{\left<#1\vphantom{#2}\right|\left.#2\vphantom{#1}\right>}
\else
\newcommand{\braket}[2]{\left<#1\vphantom{#2}\right|\left.#2\vphantom{#1}\right>}
\fi

\ifdefined\ketbra
\renewcommand{\ketbra}[2]{\left|#1\vphantom{#2}\right>\left<#2\vphantom{#1}\right|}
\else
\newcommand{\ketbra}[2]{\left|#1\vphantom{#2}\right>\left<#2\vphantom{#1}\right|}
\fi

\ifdefined\scalprod
\renewcommand{\scalprod}[2]{\left(#1\vphantom{#2}\right|\left.#2\vphantom{#1}\right)}
\else
\newcommand{\scalprod}[2]{\left(#1\vphantom{#2}\right|\left.#2\vphantom{#1}\right)}
\fi

\ifdefined\fixmatrix
\renewcommand{\fixmatrix}[2]{\left(\begin{array}{*{9}{@{}>{\centering\arraybackslash $}m{#1}<{$ }@{}}}#2\end{array}\right)}
\else
\newcommand{\fixmatrix}[2]{\left(\begin{array}{*{9}{@{}>{\centering\arraybackslash $}m{#1}<{$ }@{}}}#2\end{array}\right)}
\fi

\ifdefined\fixgausselim
\renewcommand{\fixgausselim}[4]{\left(\hspace{-1mm}\begin{array}{*{9}{@{}>{\centering\arraybackslash $}m{#1}<{$ }@{}}}#3\end{array}\vphantom{\begin{array}{*{100}c}#4\end{array}}\hspace{-1mm}\right|\hspace{-1mm}\left.\begin{array}{*{9}{@{}>{\centering\arraybackslash $}m{#2}<{$ }@{}}}#4\end{array}\vphantom{\begin{array}{*{100}c}#3\end{array}}\right)}
\else
\newcommand{\fixgausselim}[4]{\left(\hspace{-1mm}\begin{array}{*{9}{@{}>{\centering\arraybackslash $}m{#1}<{$ }@{}}}#3\end{array}\vphantom{\begin{array}{*{100}c}#4\end{array}}\hspace{-1mm}\right|\hspace{-1mm}\left.\begin{array}{*{9}{@{}>{\centering\arraybackslash $}m{#2}<{$ }@{}}}#4\end{array}\vphantom{\begin{array}{*{100}c}#3\end{array}}\right)}
\fi

\ifdefined\gausselim
\renewcommand{\gausselim}[2]{\left(\begin{matrix}#1\end{matrix}\vphantom{\begin{matrix}#2\end{matrix}}\hspace{1mm}\right|\left.\begin{matrix}#2\end{matrix}\vphantom{\begin{matrix}#1\end{matrix}}\right)}
\else
\newcommand{\gausselim}[2]{\left(\begin{matrix}#1\end{matrix}\vphantom{\begin{matrix}#2\end{matrix}}\hspace{1mm}\right|\left.\begin{matrix}#2\end{matrix}\vphantom{\begin{matrix}#1\end{matrix}}\right)}
\fi

\ifdefined\matrixel
\renewcommand{\matrixel}[3]{\left<#1\vphantom{#2#3}\right|#2\left|#3\vphantom{#1#2}\right>} 
\else
\newcommand{\matrixel}[3]{\left<#1\vphantom{#2#3}\right|#2\left|#3\vphantom{#1#2}\right>} 
\fi

\ifdefined\contravcov
\renewcommand{\contravcov}[3]{{{#1}^{#2}_{}}_{#3}}
\else
\newcommand{\contravcov}[3]{{{#1}^{#2}_{}}_{#3}}
\fi

\ifdefined\covcontrav
\renewcommand{\covcontrav}[3]{{{#1}_{#2}^{}}^{#3}}
\else
\newcommand{\covcontrav}[3]{{{#1}_{#2}^{}}^{#3}}
\fi

\ifdefined\am
\renewcommand{\am}{{\hat{a}^{\vphantom\dagger}}}
\else
\newcommand{\am}{{\hat{a}^{\vphantom\dagger}}}
\fi

\ifdefined\ap
\renewcommand{\ap}{{\hat{a}^\dagger}}
\else
\newcommand{\ap}{{\hat{a}^\dagger}}
\fi

\ifdefined\bm
\renewcommand{\bm}{{\hat{b}^{\vphantom\dagger}}}
\else
\newcommand{\bm}{{\hat{b}^{\vphantom\dagger}}}
\fi

\ifdefined\bp
\renewcommand{\bp}{{\hat{b}^\dagger}}
\else
\newcommand{\bp}{{\hat{b}^\dagger}}
\fi

\ifdefined\arctg
\renewcommand{\arctg}{\operatorname{arctg}}
\else
\newcommand{\arctg}{\operatorname{arctg}}
\fi

\usepackage{lmodern,graphics,graphicx,amsmath,amssymb,array,hhline,color,float,cite,enumitem,url}
\usepackage[left=2cm,right=2cm,top=3cm,bottom=3cm]{geometry}
\usepackage[utf8]{inputenc}
\usepackage[magyar,english]{babel}

\begin{document}
\title{Coulomb final state interaction in heavy ion collisions for L\'evy sources}
\author{M\'at\'e Csan\'ad$^{1}$, S\'andor L\"ok\"os$^{1,2}$ and M\'arton Nagy $^{1}$\\
$^{1}$ Eötvös Loránd University, H-1111 Budapest, P\'azm\'any P\'eter s\'et\'any 1/A\\
$^{2}$ Eszterh\'azy K\'aroly University, H-3200 Gyöngyös, Mátrai út 36.}

\maketitle

\abstract{Investigation of momentum space correlations of particles produced in high energy reactions requires taking final state interactions into account,
a crucial point of any such analysis. Coulomb interaction between charged particles is the most important such effect. In small systems like those created
in $e^{+}e^\textmd{--}$ or p+p collisions, the so-called Gamow factor (valid for a point-like particle source) gives an acceptable description of the
Coulomb interaction. However, in larger systems such as central or mid-central heavy ion collisions, more involved approaches are needed.
In this paper we investigate the Coulomb final state interaction for L\'evy-type source functions that were recently shown to be of much interest
for a refined description of the space-time picture of particle production in heavy-ion collisions.}

\section{Introduction}

Coulomb repulsion is the most important final state interaction that has to be considered in Bose-Einstein correlation measurements in high-energy physics.
In $e^{+}e^\textmd{--}$ or p+p collisions, where the particle emitting source is much smaller than the wavelength corresponding to the relative momentum
of the particle pair, the well-known Gamow factor (essentially the value of the Coulomb interacting pair wave function at the origin) can be used to
``correct'' for the Coulomb effect. However, for an extended source the Gamow factor overestimates the correction. A more advanced approach is to take
the source-averaged Coulomb wave function (instead of its value at the origin, which may be valid then for a point-like source), see
e.g.~Refs.~\cite{Bowler:1991vx,Sinyukov:1998fc}. In these papers a method (aptly referred to as the Bowler--Sinyukov method) is also described that
is widely used to take the effect of long-lived resonances into account.

Traditionally one assumes simple source function shapes (such as exponential, Gaussian ones) for calculating the source averaged Coulomb wave function
(as e.g. in the papers referred above); we may mention that more general sources are also considered e.g.~in Ref.~\cite{Biyajima:1995ig}.
Recently, an even more general type of source functions, namely L\'evy sources~\cite{Csorgo:2003uv,Adare:2017vig} got much interest.
L\'evy-type source functions simplify to Cauchy as well as to Gaussian ones in special cases, and allow for a more refined treatment of the space-time
picture of the particle emission. Moreover, a certain parameter of a L\'evy distribution (the so-called L\'evy exponent) also may carry information
about the order of the phase transition between deconfined and hadronic matter~\cite{Csorgo:2005it,Adare:2017vig}.

Our objective in this paper is to tackle the effect of Coulomb interaction for the case of L\'evy-type sources. (Because of the slow, power-law-like
decay of L\'evy-type sources at large distances, many previously developed methods are unsuitable for them.) We strive for analytical approximate
methods that are well suited for use in the actual treatment of experimental Bose-Einstein correlation functions.
 
\section{Coulomb effect in Bose-Einstein correlations:\\ basic concepts}

In this section we briefly review some notions and well-known formulas pertaining to the work presented hereafter.

In a statistical physical (specifically, hydrodynamical) description of particle production in high-energy collisions, a basic ingredient is the
(one-particle) {\it source function} (Wigner function), denoted here by $S(x,p)$. Its physical meaning is essentially that the probability of the
production of a particle in the infinitesimal phase-space neighborhood of momentum $\v p$ and point $\v r$ is proportional to
$S(\v r,\v p)d^3\v rd^3\v p$. Thus it is natural that the one-particle momentum distribution function $N_1(\v p)$ can be expressed as
\begin{equation}
N_1(\v p) = \int d^3\v r\,S(\v r,\v p),\qquad\tn{with the normalization}\quad
\int d^3\v p\,N_1(\v p) = 1.
\end{equation}
For a slight convenience we chose the normalization condition of $S(\v r,\v p)$ now so that $N_1(\v p)$ is now considered to be the
{\it probability distribution} of the momentum of the produced particles\footnote{
This normalizaton condition is of not much relevance here; one could just as well normalize $N_1(\v p)$ to $\langle n\rangle$, the
mean number of produced particles; $N_1(\v p)$ would then correspond to the real momentum space distribution function.}.

According to a simple quantum mechanical treatment of Bose-Einstein correlation effects, the two-particle momentum distribution function
$N_2(\v q,\v K)$ can be expressed~\cite{Yano:1978gk} with the source distribution function $S(\v r,\v p)$ as an integral over the two-particle final state wave function,
\begin{equation}
N_2(\v p_1,\v p_2) = \int d^3\v r_1 d^3\v r_2\,S(\v r_1,\v p_1)S(\v r_2,\v p_2)|\psi_{\v p_1,\v p_2}^{(2)}(\v r_1,\v r_2)|^2.
\end{equation}

The two-particle wave function $\psi^{(2)}$ must be {\it symmetric} in the space variables (for bosons); this is the main reason for the appearance
of quantum statistical (Bose-Einstein) correlations. 

With some trivial simplifications, we thus get the correlation function $C_2(\v q,\v K)$ as
\begin{equation}
C_2(\v q,\v K) = \int d^3\v r\,D(\v r,\v p_1,\v p_2) |\psi_{\v q}^{(2)}|^2,\,\quad\tn{where}\,\quad
D(\v r,\v p_1,\v p_2) = \int d^3\v R\,S\Big(\v R{+}\fract{\v r}2,\v p_1\Big)S\Big(\v R{-}\fract{\v r}2,\v p_2\Big).
\end{equation}

The momenta of the two particles were denoted by $\v p_1$ and $\v p_2$, and we used the combinations of these, the $\v q$ relative momentum and the
$\v K$ average momentum as
\begin{equation}
\v q = \v p_1-\v p_2,\qquad \v K = \rec 2\big(\v p_1+\v p_2\big) .
\end{equation}
The notation $D(\v r,\v p_1,\v p_2)$ was introduced for the so-called two-particle source function, obtained as indicated, by integrating over the
average spatial position $\v R$ of the particle pair (with $\v r\equiv\v r_1{-}\v r_2$ thus standing for the relative coordinate).

The (symmetrized) two-particle wave function may depend on all momentum and coordinate components; however, its modulus does not depend on the average
momentum $\v K$ or the average coordinate $\v R$ (owing to translational invariance). 

It is customarily assumed that the pair wave function changes much more rapidly as a function of $\v q$ than the two-particle source function $D(\v r,\v p_1, \v p_2)$. In this case we can approximate 
$\v p_1 \approx \v K$ and $\v p_2 \approx \v K$ in the arguments of $D$, and we get 
\begin{equation}
C_2(\v q,\v K)\approx \frac{\int d^3\v r\,D(\v r,\v K)|\psi^{(2)}_{\v q}(\v r)|^2}{\int d^3\v r\,D(\v r,\v K)},
\end{equation}
where we introduced $D(\v r,\v K)\equiv D(\v r,\v K,\v K) \approx  D(\v r,\v p_1,\v p_2)$.  In the special case of no final state interactions (ie.\@ when the $\psi^{(2)}$ wave function is a symmetrized plane wave), we get
the well known relation 
\begin{align}
|\psi^{(2)}_{\tn{free}}|^2 {=} 1{+}\cos(\v q\v r)\follows
C_2^{(0)}(\v q,\v  K)\approx 1 + \frac{|\tilde S(\v q,\v K)|^2}{|\tilde S(\v 0,\v K)|^2},\quad\tn{with}\quad
\tilde S(\v q,\v K) = \int d^3\v r\,S(\v r,\v K)e^{i\v q\v r},
\label{eq:pure_C}
\end{align}
thus $\tilde S$ being the Fourier transform of the source function. In this formula the ${(0)}$ superscript denotes the neglection of final state
interactions.

Returning to the general, interacting case, if one assumes (according to the core-halo model, see Ref.~\cite{Csorgo:1994in}) that a certain fraction
of the particle production (denoted by $\sqrt\lambda$) happens in a narrow, few fm diameter region (,,core''), and the rest from the decay of long-lived
resonances (the contribution of which comes from a much wider region), then one can write the source as  
\begin{equation}
S(\v r,\v p) = \sqrt\lambda S_c(\v r,\v p) + (1{-}\sqrt\lambda)S_h^{R_h}(\v r,\v p),
\end{equation}
with a normalization that respects the requirement that $\lambda$ determines the relative weight of the two components:
\begin{equation}
\int d\v r\,S(\v r,\v p) = 
\int d\v r\,S_c(\v r,\v p) = 
\int d\v r\,S_h(\v r,\v p) = 1. 
\end{equation}
Here the indices $c$ and $h$ stand for core and halo, respectively. The $R_h$ ,,radius'' parameter (the characteristic size of the halo part)
will be assumed to be much higher than the experimentally resolvable distance, $r_\tn{max} \approx \hbar/Q_\tn{min}$, where $Q_\tn{min}\approx 1-2$ MeV,
the minimal mometum difference that can be resolved experimentally. 

One can also introduce the core-core, core-halo and halo-halo two-particle source functions as
\begin{align}
D(\v r,\v K) = \lambda D_{cc}(\v r,\v K) + 2\sqrt\lambda(1{-}\sqrt\lambda)D_{ch}(\v r,\v K) + (1{-}\sqrt\lambda)^2D_{hh}(\v r,\v K),
\end{align}
where the following obvious definitions were used:
\begin{equation}
D_{AB}(\v r,\v K) \equiv \int d^3\v R\,S_A\Big(\v R{+}\fract{\v r}2,\v K\Big)S_B\Big(\v R{-}\fract{\v r}2,\v K\Big),\quad\tn{for $A,B = c$ or $h$}.
\end{equation}
With a slightly yet another notation we can write the terms of $D(\v r,\v K)$ as
\begin{align}
D(\v r,\v K) = \lambda D_{cc}(\v r,\v K) + (1{-}\lambda)D_{(h)}(\v r,\v K),\quad\tn{with}\quad D_{(h)} = \frac{2\sqrt\lambda(1{-}\sqrt\lambda)D_{ch} + (1{-}\sqrt\lambda)^2D_{hh}}{1{-}\lambda},
\end{align}
where thus the $D_{(h)}$ term contains all the halo contributions, and $D_{cc}$ is just the core-core component.
Perhaps it is useful to explicitly state the (evident) normalization conditions of all these two-particle functions:
\begin{equation}
\int D(\v r,\v K)d^3\v r  = 
\int D_{cc}(\v r,\v K)d^3\v r  = 
\int D_{ch}(\v r,\v K)d^3\v r  = 
\int D_{hh}(\v r,\v K)d^3\v r  = 
\int D_{(h)}(\v r,\v K) d^3\v r = 1. 
\end{equation}

Using these definitions, the correlation function can be expressed as 
\begin{equation}
C_2(\v q,\v K)\approx \lambda \int d^3\v r\,D_{cc}(\v r,\v K)|\psi^{(2)}_{\v q}(\v r)|^2 + 
(1-\lambda)\int d^3\v r\,D_{(h)}(\v r,\v K)|\psi^{(2)}_{\v q}(\v r)|^2 . 
\end{equation}

By taking the $R_h\to\infty$ limit in the second term\footnote{Mathematically, this is the formulation of the condition that the momentum
differences corresponding to the halo size, $\hbar/R_h$ are not resolvable by any experimental apparatus. With a re-scaling of the integral by
$\v r\to R_h\v r$, taking advantage of the fact that for large distances, $\psi(\v r)$ asymptotically becomes the free plane-wave function,
one can then use Lebesgue's dominated convergence theorem on the interchangeability of integrals and limits to infer that the second integral
indeed gives 1 in the $R_h\to\infty$ limit.}, one arrives at the well-known Bowler-Sinyukov formula~\cite{Bowler:1991vx,Sinyukov:1998fc} as
\begin{equation}
\label{e:BSiny}
C_2(\v q,\v K) = 1 - \lambda + \lambda \int d^3\v r\,D_{cc}(\v r,\v K)|\psi^{(2)}_{\v q}(\v r)|^2.
\end{equation}
Specifically, in the free case (with plane-wave wave functions) one arrives at the
\begin{equation}
\label{e:BSinyfree}
C_2^{(0)}(\v q,\v K) = 1 + \lambda\frac{|\tilde S_c(\v q,\v K)|^2}{|\tilde S_c(\v 0,\v K)|^2}
\end{equation}
formula (including the normalization term, which is unity in this paper).
The experimental observation is that --- although the free correlation function defined in Eq.~\eqref{eq:pure_C} takes the
value of 2 at 0 relative momentum: $C_2^{(0)}(\textbf{0},\textbf{K})=2$, --- the measured value is $1+\lambda$. The core-halo model thus naturally
explains this fact in terms of the finite momentum resolution of any experiment. In the core-halo model the intercept of the real,
measurable correlation function at $\v q=0$ thus tells the fraction of pions coming from the core. In the Coulomb interacting
(realistic) case, the interpretation of $\lambda$ as any intercept parameter is not so simple, however. The Bowler-Sinyukov
method, Eq.~(\ref{e:BSinyfree}) gives a means to take the core-halo model into account when treating the Coulomb effect.

To investigate the $\lambda$ parameter (which, as it is directly connected to the proportion of resonance decay particles, may have interesting
physical consequences, see e.g.~Refs.~\cite{Vertesi:2009wf,Adare:2017vig}) one needs a firm grasp on the effect of final state interactions in Bose-Einstein
correlation functions. For the most important such effect, the Coulomb effect, the $\psi_{\v q}^{(2)}(\v r)$ wave
function (the two-body scattering solution of the Schr\"odinger equation with Coulomb repulsion) is well known in the center-of-mass system
of the outgoing particles (the so-called PCMS system). Its expression is 
\begin{equation}
\psi_{\v q}^{(2)}(\v r) = \rec{\sqrt 2}\frac{\Gamma(1{+}i\eta)}{e^{\pi\eta/2}}\kz{e^{i\v k\v r}F\big({-}i\eta,1,i(kr{-}\v k\v r)\big)
+ [\v r\leftrightarrow -\v r]},\quad\tn{where}\quad\v k=\frac{\v q}2.
\end{equation}
Here $F(\cdot,\cdot,\cdot)$ is the confluent hypergeometric function, $\Gamma(\cdot)$ is the Gamma function, and 
\begin{align}
\eta=\frac{q_e^2}{4\pi\varepsilon_0}\rec{\hbar c}\frac{m_\pi c^2}{q c}=\alpha_{\rm EM}\frac{m_\pi c}{q}
\end{align}
is the Sommerfeld parameter, with $q_e^2/(4\pi\varepsilon_0)$ being the Coulomb-constant, $\alpha_{\rm EM}$ the fine-stucture constant of the electromagnetic interaction, and $m_\pi$ the pion mass (as from now on, we restrict this analysis to pion pairs).

For a given source function $S(\v r,\v K)$, the ratio of the (measurable) correlation function $C_2(\v q)$ and the $C^{(0)}_2(\v q)$ function is usually
called the {\it Coulomb correction}\footnote{The terminology is not uniform here; it is sometimes this factor, and sometimes its inverse what is called
the Coulomb correction.}, $K(\v q)$:
\begin{equation}
K(\v q) = \frac{C_2(\v q)}{C^{(0)}_2(\v q)}\Follows
C^{(0)}_2(\v q) = C_2(\v q)\cdot\rec{K(\v q)} .
\end{equation}
If one focuses on the simple property of the $C^{(0)}_2(\v q)$ function as being the Fourier transform of the source, then one might want to recover
$C^{(0)}_2(\v q)$ from the measured $C_2(\v q)$: for this, one uses the Coulomb correction factor. Indeed, many assumptions have been used to estimate
the $K(\v q)$ factor: the simplest case is the so-called Gamow factor that treats the source as a point-like one when calculating $K(\v q)$:
\begin{equation}
S(\v r) = \delta^{(3)}(\v r)\Follows K(\v q) = K_\tn{Gamow}(q) = |\psi_{\v q}^{(2)}(0)|^2 = \frac{2\pi\eta}{e^{2\pi\eta}{-}1} .
\end{equation}

A method that suits the scope of heavy-ion collisions a little more would be to pre-calculate $K(\v q)$ for a single specific given assumption for $S(\v r)$,
then apply this correction (with the Bowler-Sinyukov method) and find the $S(\v r)$ from a fit to the Fourier transform of the recovered $C^{(0)}(\v q)$.
However, it is clear that this process should be done iteratively: after the first ``round'' of such fits, one would have to re-calculate the
Coulomb correction. When this iteration converges, one in principle arrives at the proper $S(\v r)$.

\section{Numerical table for the Coulomb correction for L\'evy source}

Recent studies have shown that the assumption of a L\'evy-type of source function is well suited for the description of two-particle Bose-Einstein
correlation functions. The details of the validity of the L\'evy-shape assumption is exhaustively expounded in Refs.~\cite{Csorgo:2003uv,Adare:2017vig}.
The (spherically symmetric) L\'evy distribution utilized here has two parameters, scale parameter (radius) $R$ and L\'evy index $\alpha$, and is expressed as 
\begin{align}
\c L(\alpha,R,\v r) := \int\frac{d^3\v q}{(2\pi)^3}e^{i\v q\v r}\exp\Big({-}\rect2|\v q^2R^2|^{\alpha/2}\Big).
\end{align}
In the $\alpha{=}2$ case one gets a Gaussian distribution, in the $\alpha{=}1$ case the Cauchy distribution is recovered. For other $\alpha$ values, no
simple analytic expression exists for the result of this Fourier transform-like integral. As a remark, we note that the concept of this symmetric L\'evy
distribution can be generalized without much effort to the non-spherically symmetric case by replacing $R^2$ with a symmetric $3{\times}3$
matrix $R^2_{kl}$.

In order to apply L\'evy-type sources in a self-consistent way, the Coulomb integral defined in Eq.~\eqref{e:BSiny} has to be calculated. This cannot be
carried out in a straightforward analytic manner. In the following we demonstrate two approaches that can be employed to handle the Coulomb final state
effect in the presence of a L\'evy source.

The integral in Eq.~\eqref{e:BSiny} cannot be evaluate analytically for a L\'evy source so it has to be calculated numerically. For experimental
purposes, the results can be loaded to a binary file as a lookup table and can be used in the fitting procedure (thus circumventing the need for an
iterative process for the Coulomb correction). Interpolation also should be applied since the correlation function only can be filled into the lookup
table for discrete values of the parameters. This interpolation, however, could cause numerical fluctuations in the $\chi^2$ landscape and could mislead
the fit algorithm, so an iterative procedure should be applied in the following manner:

\begin{enumerate}[leftmargin=*,labelsep=4mm]
\item	Fit with the function defined in Eq. \eqref{e:BSiny} $\Rightarrow$ $\alpha_0, R_0, \lambda_0$
\item	Fit with $C_2^{(0)}(\lambda,R,\alpha;Q)\frac{C_2(\lambda_0,R_0,\alpha_0;Q)}{C_2^{(0)}(\lambda_0,R_0,\alpha_0;Q)}$ $\Rightarrow$ $\lambda_1, R_1, \alpha_1$
\item	Repeat while $\lambda_1, R_1, \alpha_1$ and $\lambda_0, R_0, \alpha_0$ differ less then 1\%
\end{enumerate}
In this manner, the fit parameters $\lambda(K), R(K), \alpha(K)$ can be yielded. This technique was used in Ref. \cite{Adare:2017vig}.

\section{Parametrization of the Coulomb correction for L\'evy source}

In this section, let us review a different approach, where based on the numerical table mentioned above, a parametrization can be formulated.
In other words, one can get the Coulomb correction values from the table and parametrize its
$R$ and $\alpha$ dependences. This approach was encouraged by the successful parametrization of the $\alpha{=}1$ case
(the Cauchy case) done by the CMS collaboration (see Ref.~\cite{Sirunyan:2017ies}, Eq. (5) for details). This can be considered as our starting
point for the more general, L\'evy case (for arbitrary $\alpha$). The expression used by CMS for the Cauchy distribution, $\alpha{=}1$ was
\begin{equation}
K(\v q)_\tn{Cauchy} = K_\tn{Gamow}(\v q)\times\z{1+\frac{\alpha_\tn{EM}\pi m_\pi R}{1.26\hbar c + qR}},\quad\tn{where}\quad
\alpha_\tn{EM} = \frac{q_e^2}{4\pi\varepsilon_0}\rec{\hbar c}\approx\rec{137}.
\label{eq:Sikler}
\end{equation}
Generally, this is a correction of the Gamow correction. This simple formula has the advantage of having only 1 numerical constant parameter (the
1.26 in the denominator). However, it assumes $\alpha{=}1$, and we look for a generalization for arbitrary L\'evy $\alpha$ values.

\begin{figure}
\centering
\includegraphics[width=0.65\textwidth]{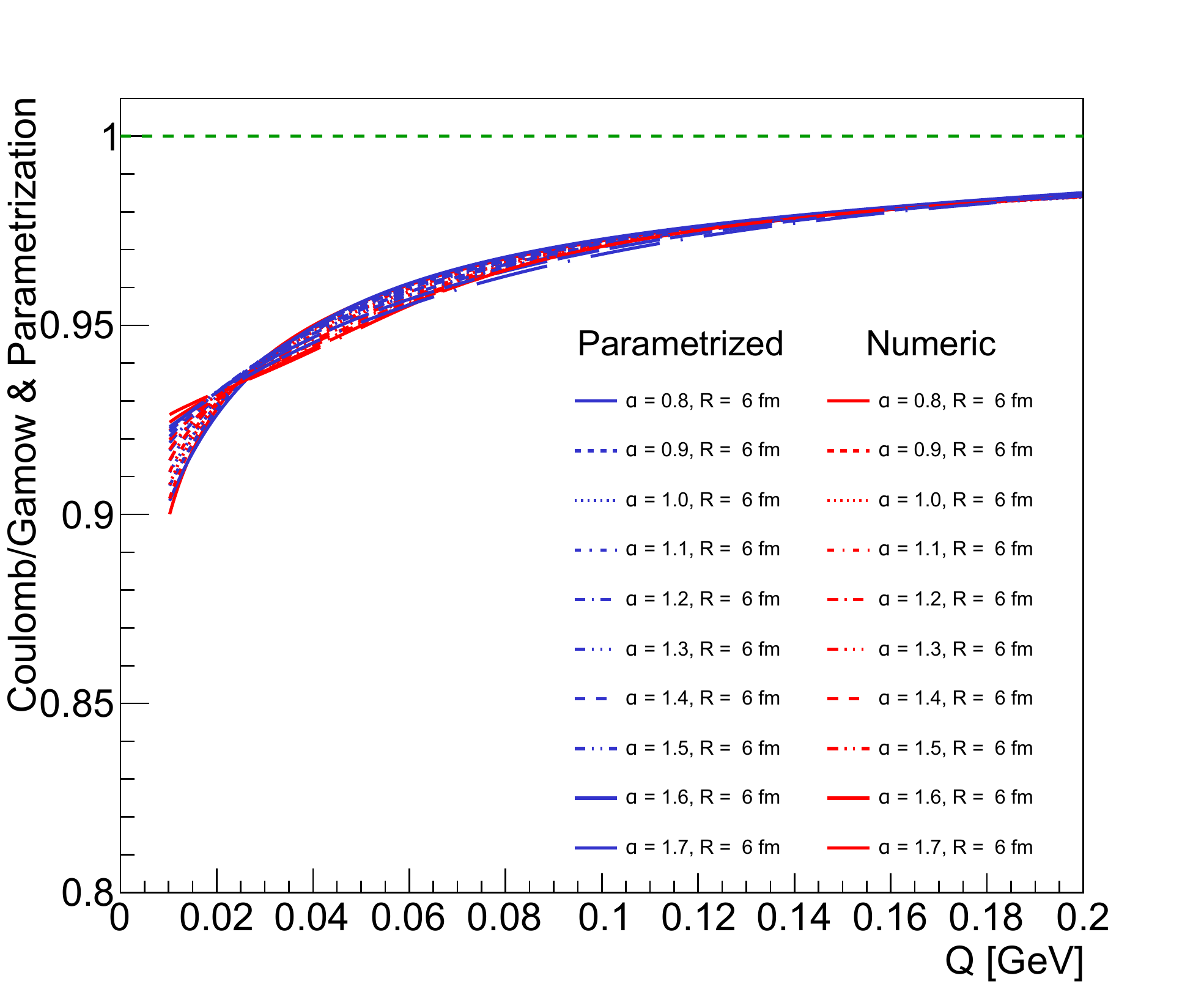}
\caption{\label{fig:basic} An example for the parametrization of the Coulomb correction, divided by the Gamow factor, for a given $R$ and a set of $\alpha$ values. One can observe that the $\alpha$ dependence is quite weak but still observable and quite complicated.}
\end{figure}

A more general correction for the Gamow correction which is able to describe the Coulomb correction for a L\'evy source has to fulfill the following
requirements:
\begin{itemize}
	\item It should follow not only the $R$, but the $\alpha$ dependence.
	\item In $\alpha{=}1$ case, it should reduce to Eq.~\eqref{eq:Sikler}.
\end{itemize}
To fulfill these, we replace $R$ with $R/\alpha$ to introduce the $\alpha$-dependence and take higher order terms in
$\frac{q R}{\alpha\hbar c}$ into consideration. Our trial formula is then assumed to be
\begin{align}
K_\tn{L\'evy}(q,\alpha,R) &= K_\tn{Gamow}(\v q)\times K_\tn{mod}(\v q),\quad\tn{with}\nonumber\\
K_\tn{mod}(\v q) &= 1 +  \frac{A(\alpha,R)\frac{\alpha_\tn{EM}\pi m_\pi R}{\alpha \hbar c}}
{1+B(\alpha,R)\frac{qR}{\alpha \hbar c}+C(\alpha,R)\z{\frac{q R}{\alpha \hbar c}}^2+D(\alpha,R)\z{\frac{q R}{\alpha \hbar c}}^4} .\label{e:KLevyparam}
\end{align}
and the task is to find a suitable choice for the $A(\alpha,R)$, $B(\alpha,R)$, $C(\alpha,R)$, $D(\alpha,R)$ functions that yield an acceptable
approximation of the results of the numerical integration (contained in our lookup table). The assumed form seems to be sufficient since it
simplifies to Eq.~\eqref{eq:Sikler} if $\alpha{=}1$ and $C{=}D{=}0$, and could follow the observed weak $\alpha$ dependence of the Coulomb integral
(see Fig.~\ref{fig:basic}).

We fitted the above (\ref{e:KLevyparam}) formula to the numerically calculated results for $\alpha$ parameter values between 0.8 and 1.7 and $R$ parameter values between 3 fm and 12 fm (the ranges were motivated by the PHENIX results of Ref.~\cite{Adare:2017vig}). With this we obtained the $A$, $B$, $C$, $D$ values as a function of the given $\alpha$ and $R$ parameters. As a next step, we also parametrized these dependencies empirically, and found that the following expressions give satisfactory agreement with the lookup table:
\begin{align}
A(\alpha,R) &=  (a_A\alpha + a_B)^2 + (a_C R + a_D)^2 + a_E (\alpha R + 1)^2 \\
B(\alpha,R) &=  \frac{ 1 + b_A R^{b_B} - \alpha^{b_C}}{ \alpha^{2} R ( \alpha^{b_D} + b_E R^{b_F} ) } \\
C(\alpha,R) &=  \frac{ c_A + \alpha^{c_B} +c_C R^{c_D} }{c_E} \left( \frac{\alpha}{R} \right)^{c_F}   \\
D(\alpha,R) &=  d_A + \frac{R^{d_B} + d_C \alpha^{d_F}}{R^{d_D}\alpha^{d_E}}.
\end{align}
The parameters in these functions turn out best to have the values as follows:

\begin{table}[h!]
\centering
\begin{tabular}{c c c c c c}
$a_A$ =  0.36060 & $a_B$ = -0.54508 & $a_C$ =  0.03475 & $a_D$ = -1.30389 & $a_E$ =  0.00378 & \\
$b_A$ =  2.04017 & $b_B$ =  0.55972 & $b_C$ =  2.47224 & $b_D$ = -1.26815 & $b_E$ = -0.11767 & $b_F$ = 0.52738 \\
$c_A$ = -1.00015 & $c_B$ =  0.00012 & $c_C$ =  0.00008 & $c_D$ =  0.26986 & $c_E$ =  0.00003 & $c_F$ = 1.75202 \\
$d_A$ =  0.00263 & $d_B$ = -0.13124 & $d_C$ = -0.83149 & $d_D$ =  1.57528 & $d_E$ =  0.27568 & $d_F$ = 0.04937
\end{tabular}
\end{table}
This parametrization describes the $R$ and $\alpha$ dependence of the Coulomb integral in a range where the Coulomb correction deviates from 1 by more
than a factor of $\sim 10^{-4} - 10^{-5}$. We find that this region is 0 GeV/$c$ $<q<$ 0.2 GeV/$c$. As an example, for $R=6$ fm and with different
$\alpha$ values, we plotted the results of the parametrization on Fig. \ref{fig:basic}.

It turns out that the functional form specified above does yield a satisfactory fit at lower values of $q$, below $0.1-0.2$ GeV/$c$. However, at higher values, the fit that
is acceptable at low $q$, inevitably starts to deviate from the desired values, i.e. cannot be used to extrapolate beyond the fitted $q$ range.
The intermediate $q$ region above and around 0.1 GeV/$c$ can instead be described with an exponential-type function parametrized based on intermediate $q$ fits to the numerical
table, with the following functional form:
\begin{equation}
E(q) = 1 + A(\alpha,R)\exp\kz{ - B(\alpha,R)q},
\end{equation}
where the $A(\alpha,R)$ and $B(\alpha,R)$ functions have a form as

\begin{align}
A(\alpha,R) &= A_a + A_b\alpha + A_c R + A_d\alpha R + A_e R^2 + A_f(\alpha R)^2,\\
B(\alpha,R) &= B_a + B_b\alpha + B_c R + B_d\alpha R + B_e R^2 + B_f(\alpha R)^2.
\end{align}
The parameters were chosen based on a fit to numerically calculated Coulomb correction values, and the optimal case was found to be represented by these parameter values:

\begin{table}[h!]
\centering
\begin{tabular}{c c c c c c}
$A_a$ =  0.20737 & $A_b$ = -0.00999 & $A_c$ = -0.02671 & $A_d$ = -0.00373 & $A_e$ =  0.00119 & $A_f$ =  0.00016 \\
$B_a$ = 25.80500 & $B_b$ =  4.01674 & $B_c$ =  0.00873 & $B_d$ = -0.25606 & $B_e$ =  0.01077 & $B_f$ = -0.00270
\end{tabular}
\end{table}

This exponential damping factor is ``joined'' to the proper parametrization valid for the interesting $q$ range by a Wood-Saxon-type of cut-off function:
\begin{equation}
F(q) = \rec{1{+}\exp\z{\frac{q-q_0}{D_q}}},
\end{equation}
where $q_0{=}0.08$ GeV and $D_q {=} 0.02$ GeV. We investigated different cut-off functions, such as $1/(1+(q/q_0)^{n})$, but found that the results are rather independent from this choice.

\begin{figure}
\centering
\includegraphics[width=0.7\textwidth]{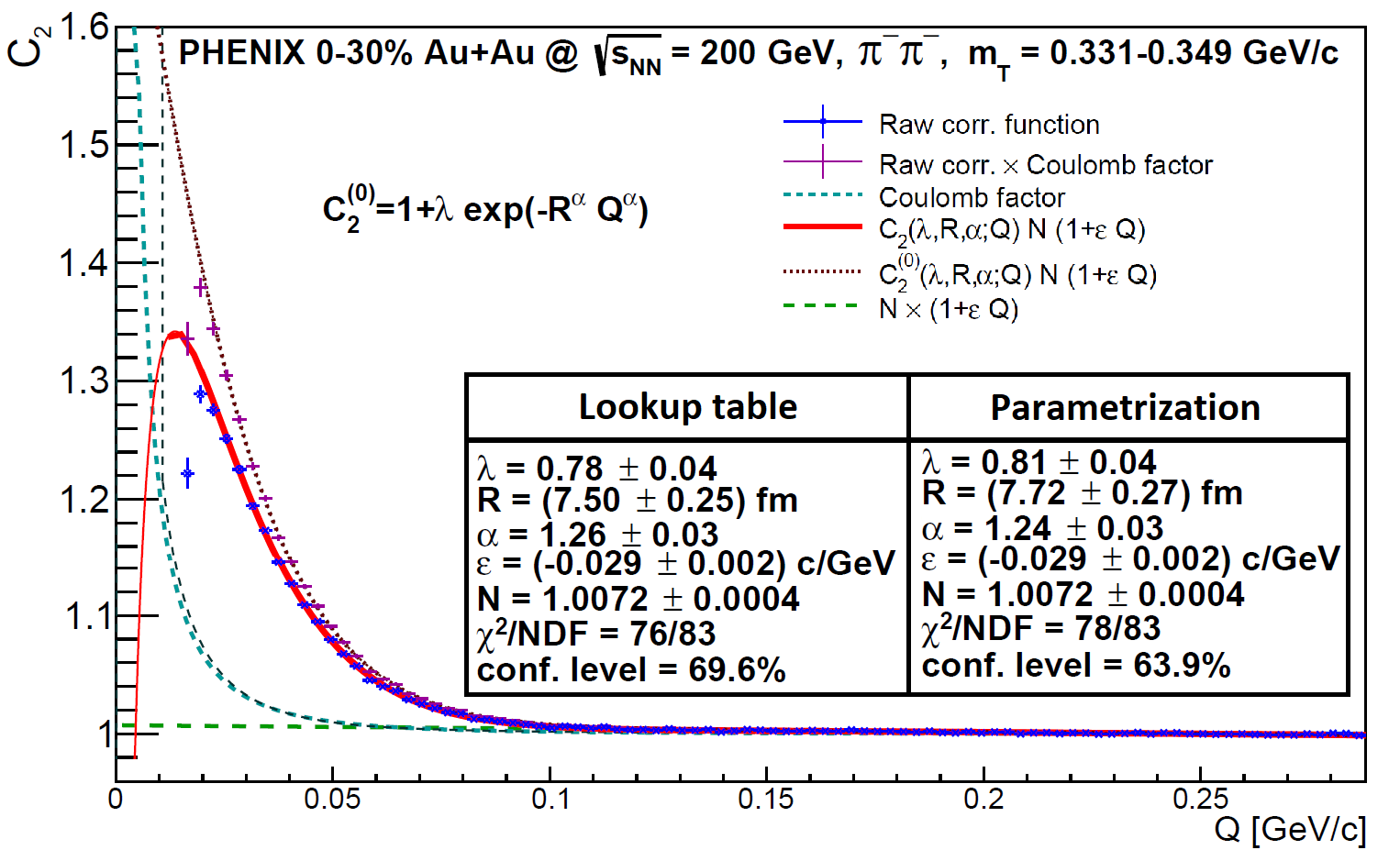}
\caption{\label{fig:repord} The reproduction of earlier PHENIX results~\cite{Adare:2017vig} with the parametrization. The original PHENIX
fit procedure employed the lookup numerical table, here we show our results from the parameterization.}
\end{figure}

\begin{figure}
\centering
\includegraphics[width=0.85\textwidth]{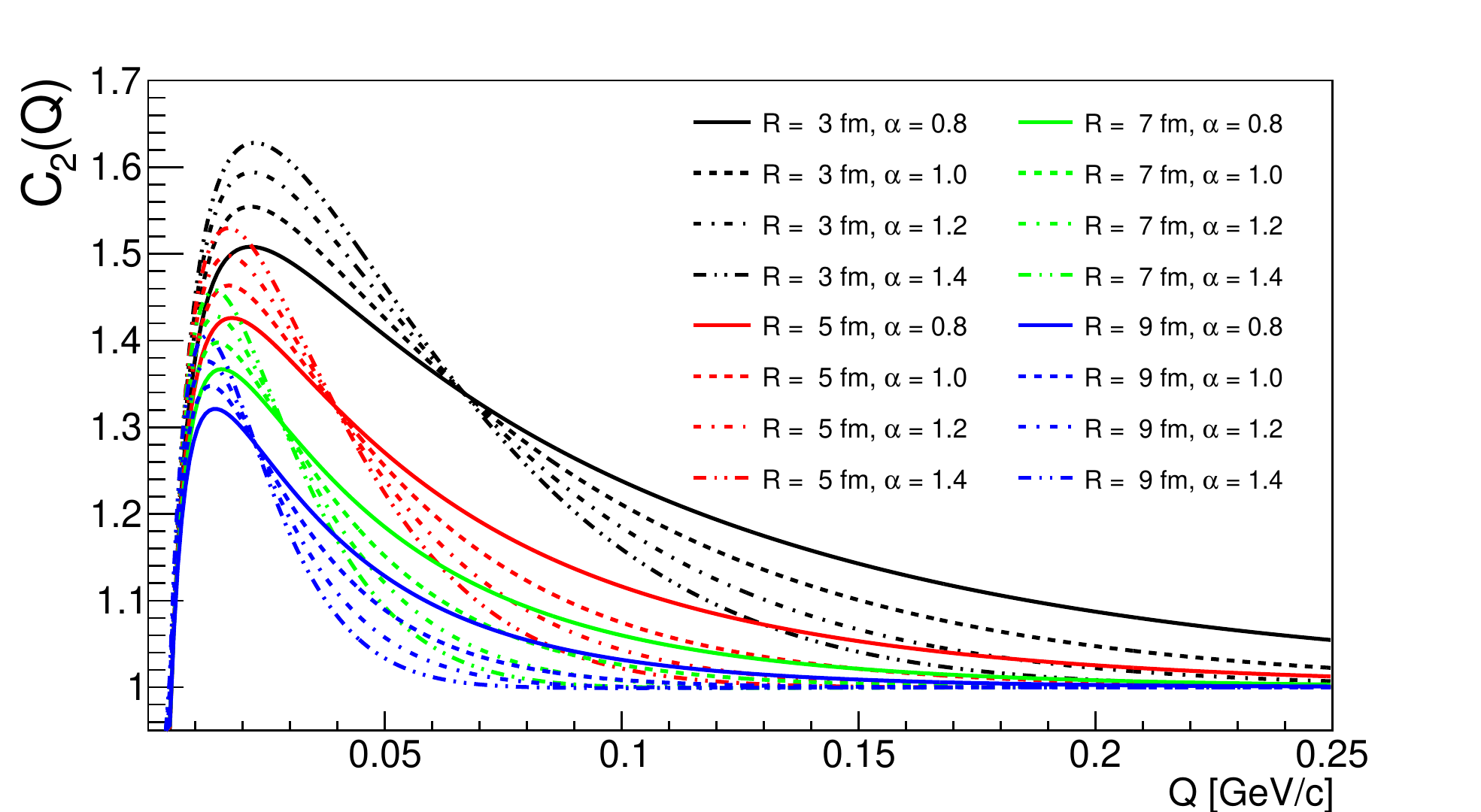}
\caption{\label{fig:C2example} Example correlation functions, based on Eq.~(\ref{e:C2full}), for different values of parameters $R$ and $\alpha$.}
\end{figure}

Putting all of the above together, our final parametrization, valid for $\alpha=0.8-1.7$ and $R=3-12$ fm values, is thus
\begin{equation}
K(q,\alpha,R)^{-1} =  F(q)\times K_\tn{Gamow}^{-1}(q)\times K_\tn{mod}^{-1}(q;\alpha,R) + (1{-}F(q))\times E(q)\label{e:coulcorrparam}
\end{equation}
and the Coulomb corrected correlation function which could be fitted to data, can be written in the form of
\begin{equation}
C_2(q;\alpha,R) = \left[ 1 - \lambda + K(q;\alpha,R) \lambda \left( 1 + \exp \left[ | q R |^{\alpha} \right] \right) \right] \cdot (\textmd{assumed background}).\label{e:C2full}
\end{equation}
We used this formula to reproduce earlier PHENIX results from Fig. 3. of Ref.~\cite{Adare:2017vig}~\footnote{The data of the shown PHENIX correlation function result was retrieved from \url{https://www.phenix.bnl.gov/phenix/WWW/info/data/ppg194_data.html}.}; this can be seen on Fig.~\ref{fig:repord}. The two fits are compatible with each other.
For an example code calculating the formula of (\ref{e:coulcorrparam}), please see Ref.~\cite{coulcorrparamcode}. Example curves resulting from the above (\ref{e:C2full}) formula (with the background being unity) are shown in Fig.~\ref{fig:C2example}. These clearly show how $R$ changes the scale, and $\alpha$ changes the shape of the correlation functions. Parameter $\lambda$ provides an overall normalization to the distance of these curves from unity, as described by Eq.~(\ref{e:C2full}).

\begin{figure}
\centering
\includegraphics[width=0.7\textwidth]{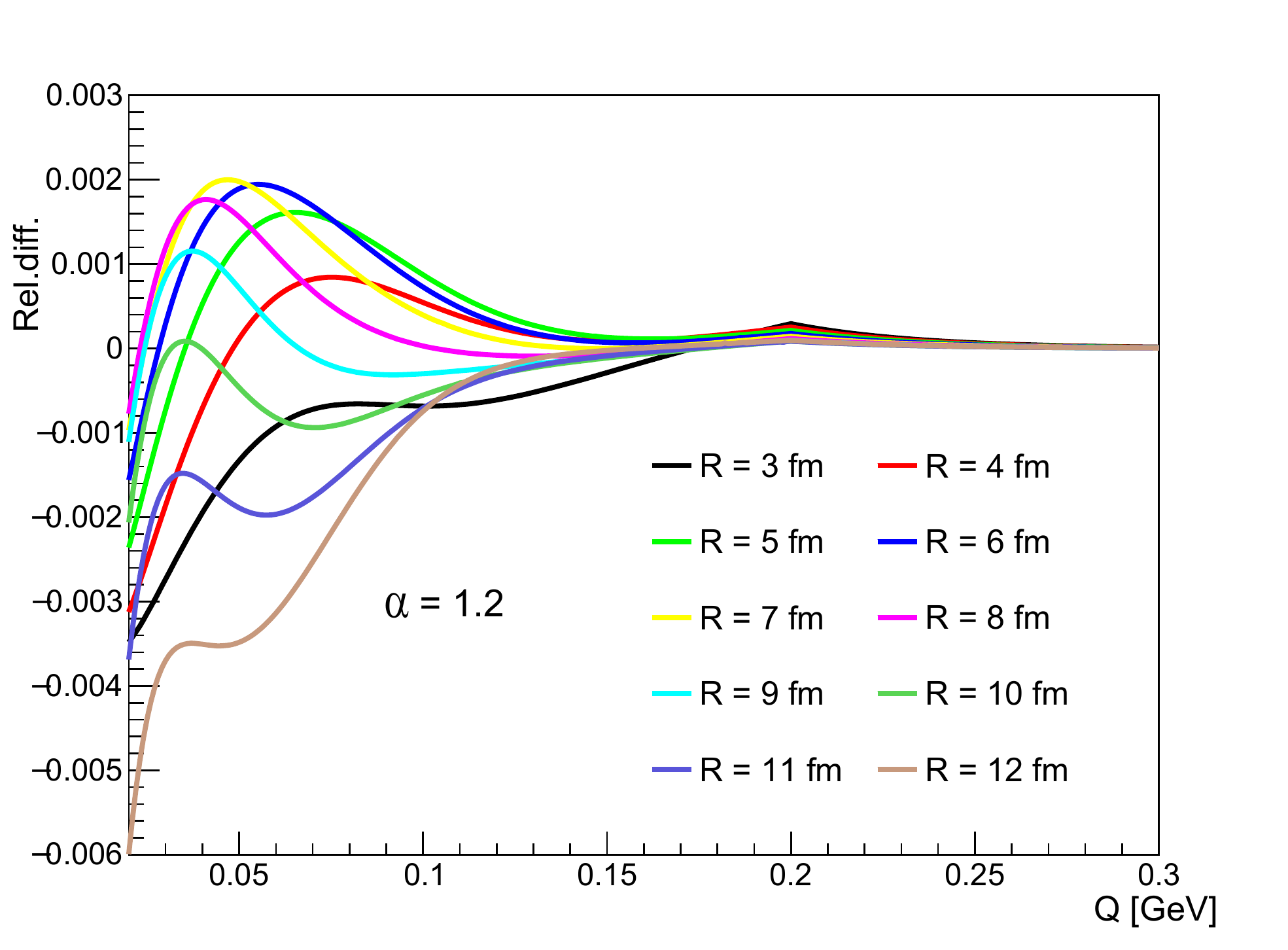}
\caption{\label{fig:reldev} The relative deviation of the parametrization measured in $\%$ form the table for a given $\alpha$ with various $R$ values within the domain of the parametrization.}
\end{figure}

\begin{figure}
\centering
\includegraphics[width=0.7\textwidth]{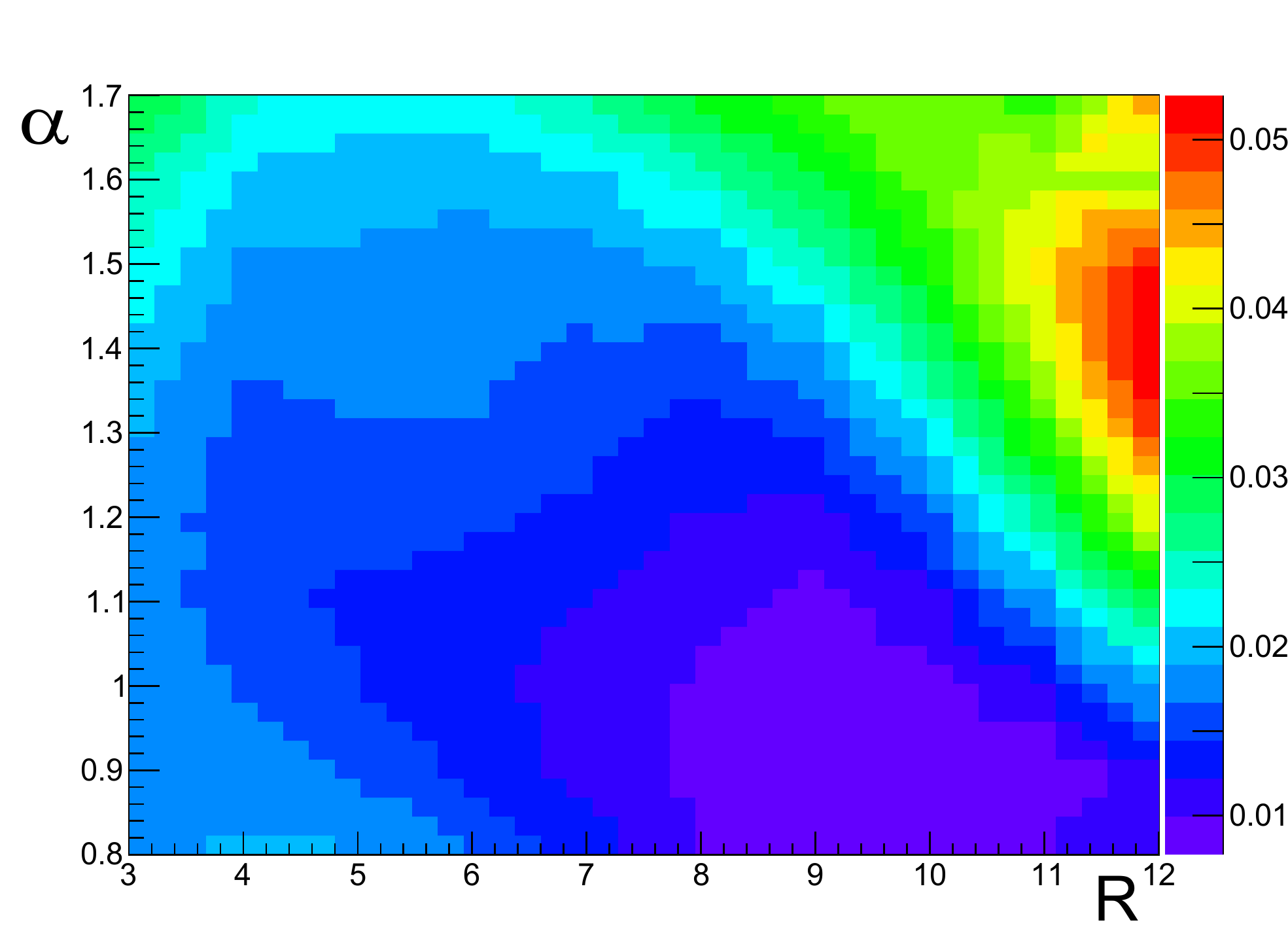}
\caption{\label{fig:reldev2D} The relative deviation of the parametrization measured in $\%$ form the table for the domain of the parameterization in $\alpha$ ($0.8-1.7$) and in $R$ ($3-12$ fm),
averaged over a $q$ region of 0.01 to 0.1 GeV$/c$.}
\end{figure}

We investigated the parametrization by means of its relative deviation from the lookup table. The results can be seen in Fig.~\ref{fig:reldev}. In the
case when $\alpha{=}1.2$ with different $R$ values, we present a two-dimensional histogram of the relative differences in in Fig.~\ref{fig:reldev2D}.
The maximum of these relative differences is around 0.05\%.

\section{Conclusions}

We investigated the Coulomb correction of Bose-Einstein correlations in high energy heavy ion reactions under the assumption of L\'evy source functions. We outlined two equivalent methods that are suited for an experimental analysis. One of them is a numerical lookup table, another one is a parametrization obtained from the former. We investigated the accuracy of the methods and found that a not very complicated ad-hoc parametrization, in the well defined parameter range of $R=3-12$ fm and $\alpha=0.8-1.7$, provides an experimentally acceptable description of the results of the numerical integration that is required for the handling of the Coulomb effect. Our parametrization can thus be used effectively in HBT correlation analyses that assume L\'evy-type source functions.


\section*{Acknowledgments}This work was supported by the NKFIH grant FK 123842. S.L. is grateful for the support of EFOP 3.6.1-16-2016-00001.
M.N. and M.Cs. are supported by the Hungarian Academy of Sciences through the ``Bolyai J\'anos'' Research Scholarship program as
well as the \'UNKP-18-4 New National Excellence Program of the Hungarian Ministry of Human Capacities.

\end{document}